\documentclass[preprint,12pt]{elsarticle}

\usepackage{graphicx}
\usepackage{overpic}
\usepackage{subcaption}
\usepackage{amssymb}
\usepackage{amsmath}
\usepackage{comment}
\usepackage{color}
\usepackage{soul}
\usepackage{booktabs}
\usepackage{tabularx}
\usepackage{siunitx}
\usepackage{algorithm}
\usepackage{algpseudocode}
\usepackage{algcompatible}
\usepackage[left=1in, right=1in]{geometry}
\usepackage{multirow}

\newcommand\Rey{\mbox{\textit{Re}}}  
\newcommand\Sto{\mbox{\textit{St}}} 
\newcommand\Cor{\mbox{\textit{COR}}}

\usepackage{setspace}

\begin{document}
\doublespacing
\begin{frontmatter}



\title{A numerical study on the coefficient of restitution of wet collisions}


\author[1]{Abhishek Kumar Singh \corref{cor1}}
\author[2]{ Christopher Robert Kit Windows-Yule}
\author[1]{Prapanch Nair}

\affiliation[1]{organization={Department of Applied Mechanics, Indian Institute of Technology Delhi},
            addressline={Hauz Khas}, 
            city={New Delhi},
            postcode={110016}, 
            country={India}}
\affiliation[2]{organization={School of Chemical Engineering, University of Birmingham},
            addressline={Edgbaston}, 
            city={Birmingham}, 
            postcode={B15 2TT},
            country={UK}}
            
\cortext[cor1]{Corresponding author: singh.abhishekmech22@gmail.com; Tel.: +91-7049256260}
\begin{abstract}
Using smoothed particle hydrodynamics (SPH) simulations, we investigate the coefficient of restitution (COR) in wet collisions and identify a scaling law governing its behavior. The simulations employ an updated-Lagrangian, mesh-free framework that is validated against experimental measurements. We neglect surface tension effects since the impact conditions correspond to a moderate-to-high Weber number regime. The COR is found to depend on the Stokes number and a dimensionless film thickness defined as the ratio of the liquid film thickness to the diameter of the impacting solid bead. Two distinct regimes are observed, each characterized by different power-law exponents.   
\end{abstract}


\end{frontmatter}

\section{Introduction}
Estimation of the energy dissipation in wet particle collisions is crucial in modelling several granular flow phenomena, including debris flows~\cite{Jop2006}, resonance acoustic mixing~(RAM)~\cite{Lewis_RAM}, wet granulation, coating, milling, and sediment transport at industrial scales~\cite{Tang2017}.
For dry collisions, the coefficient of restitution depends only weakly on impact velocity and is often treated as a material constant within kinetic-theory-based descriptions of granular flows~\cite{Brilliantov1996}. However, the presence of an interstitial liquid film fundamentally alters this picture by introducing additional dissipation mechanisms associated with viscous flow within the liquid layer, inertial pressure buildup, and deformation of the liquid free surface. Currently, most macroscopic models assume wet collisions are governed by the Stokes number alone \cite{hussain}.

Early experimental and theoretical studies established that wet collisions can dissipate significantly more energy than dry impacts. Joseph \textit{et al.}~\cite{Joseph2001} investigated collisions between rigid spheres coated with thin liquid films and demonstrated that viscous losses due to squeeze flow dominate rebound dynamics at low impact velocities. These observations are consistent with classical analyses of squeeze flow between approaching bodies in viscous fluids, which form the basis of lubrication-theory descriptions of wet impacts~\cite{Davis1986}. Such models successfully capture the very low Reynolds number regime ($\Rey<<1$), where fluid inertia and free-surface deformation may be neglected.

Merely lubrication-based descriptions fail at higher velocities or for moderately thick films, where fluid inertia and transient pressure buildup significantly influence rebound dynamics. Kantak and Davis~\cite{Kantak2013} showed that deviations from lubrication-theory predictions emerge as inertial effects become significant, even for relatively thin liquid layers.

A major experimental advance in understanding this behavior was provided by Gollwitzer \textit{et al.}~\cite{Gollwitzer}. They  systematically measured the normal impacts of glass beads onto thin liquid films and reported that the wet COR increases with impact velocity before approaching a saturation value at high velocities. Their results were presented as a power-law function of the Stokes number for a fixed liquid film-thickness to particle-diameter ratio, highlighting the influence of particle inertia relative to viscous dissipation. However, the effect of varying the liquid-film thickness to particle-diameter ratio ($\delta/D$) independently of the Stokes number was not explored experimentally.

Motivated by these findings, several studies expressed the wet COR as a power-law of the Stokes number (St) alone, implicitly assuming a single dominant dissipation mechanism~\cite{Kantak2013}. However, the experimental results of Gollwitzer \textit{et al.} ~\cite{Gollwitzer} and subsequent numerical studies~\cite{Sun2016} indicate that this assumption breaks down when $\delta/D$ varies or when inertial effects within the liquid film become important. These observations indicate that multiple dissipation mechanisms govern wet collision dynamics, with relative importance set by impact conditions and liquid-film thickness.

In the present work, we address this gap using incompressible smoothed particle hydrodynamics (ISPH) simulations~\cite{Cummins1999, nair2014improved}, a projection-based incompressible formulation within the broader SPH framework~\cite{Monaghan2005}, to study the normal impact of a rigid sphere onto a thin liquid film in the moderate-to-high Weber number regime. By systematically varying particle diameter, impact velocity, liquid viscosity, and film thickness, we identify the governing dimensionless parameters controlling rebound dynamics. 
In ~\S\ref{sec:rigidbody} we explain the collision model followed by the simulation methodology in~\S\ref{sec:methodology}. We present our simulation setup in~\S\ref{sec:setup}. Further, in~\S\ref{sec:validation} we validate the numerical simulations against experiments. In \S \ref{sec:results}, we present simulations that span both $\Sto$ and a dimensionless film thickness $\gamma$ (defined as $\delta / D$, where $\delta$ denotes the liquid-film thickness and $D$ the particle diameter) as independent non-dimensional variables.
We show two distinct regimes with different pairs of scaling exponents using these simulations. We then conclude by explaining the two regimes using kinetic energy dissipation.

\section{Methodology}
\label{sec:methodology}
\subsection{Incompressible Smoothed Particle Hydrodynamics Formulation (ISPH)}
\label{sec:isph}
An incompressible smoothed particle hydrodynamics (ISPH) formulation is adopted to resolve pressure-driven dissipation and fluid–solid interaction forces during short-duration wet impacts. Enforcing incompressibility leads to stable and physically consistent pressure fields, which are crucial for capturing the energy loss mechanisms governing the coefficient of restitution. The formulation follows the projection-based approach introduced by Cummins and Rudman~\cite{Cummins1999}, while kernel interpolation and stability considerations are based on standard SPH formulations~\cite{Monaghan2005}.

\par
The ISPH simulations employ a Cummins time integration scheme~\cite{Cummins1999} together with the Cleary--Monaghan viscous model~\cite{Cleary1999}. A smoothing length ratio of $h/\Delta x = 2$ is used. The pressure Poisson equation is solved using the BiConjugate Gradient Stabilized (BiCGSTAB) method~\cite{VanDerVorst1992} with Jacobi preconditioning and a convergence tolerance of $10^{-9}$. The time step is selected based on Courant--Friedrichs--Lewy (CFL) stability constraints accounting for both inertial and viscous effects~\cite{Monaghan1994}. The effect of gravity is considered throughout all simulations, with a constant gravitational acceleration of $9.81~\mathrm{m/s^2}$ acting in the negative vertical direction. The free surface pressure boundary condition for solving the pressure Poisson equation is applied using the method introduced in \cite{nair2014improved}.


\subsection{Rigid-body dynamics and collision modelling}
\label{sec:rigidbody}
The solid bead is modelled as a free (unconstrained) rigid body composed of a cluster of Lagrangian particles. The dynamics of the rigid body is fully coupled to the surrounding fluid through hydrodynamic forces obtained from the ISPH formulation \cite{nair2014improved, nair2017study}. 

Though the rigid body is composed of Lagrangian points, such a discretization is necessary only in regions where interaction with the fluid domain is present. Therefore, we use Lagrangian discretization of the rigid body only in a truncated region which comes in proximity (within kernel cutoff radius of the SPH particles) with the liquid film, as shown in Fig. \ref{fig:setup}. Also a solid sphere is not necessary and hence a spherical shell of thickness of 4 particle-widths is considered. This greatly reduces computational cost. Once the individual interaction forces are computed at these Lagrangian points, total force and moment on the rigid body about its \emph{true} center of gravity is computed by appropriately summing these contributions. Thereafter the rigid body dynamics is solved as described below. 

\subsubsection{Linear and Angular momentum balance}
\label{sec:momentum}
Let $\mathcal{R}$ denote the rigid body and let $A$ be its centre of gravity (COG). The total mass of the body is
\begin{equation}
m = \sum_{a\in\mathcal{R}} m_a ,
\end{equation}
where $m_a$ is the mass of particle $a$. The resultant external force acting on the rigid body is obtained by summing the forces acting on its constituent particles, yielding the translational equation of motion
\begin{equation}
\vec{F}_A = m \vec{a}_A = \sum_{a\in\mathcal{R}} m_a \vec{a}_a ,
\end{equation}
where $\vec{a}_A$ is the acceleration of the COG. Internal constraint forces between rigid-body particles cancel identically and therefore do not contribute to the net force.

The angular momentum balance is expressed about the COG as
\begin{equation}
\vec{M}_A = \sum_{a\in\mathcal{R}} \vec{r}_{Aa} \times m_a \vec{a}_a ,
\end{equation}
where $\vec{r}_{Aa}$ denotes the position vector of particle $a$ relative to the COG.


The rotational motion of the rigid body is formulated in a body-fixed reference frame aligned with the principal axes of inertia. Let $\{\vec{e}_1,\vec{e}_2,\vec{e}_3\}$ denote the orthonormal principal directions, and let $I_{11}^A$, $I_{22}^A$, and $I_{33}^A$ be the corresponding principal moments of inertia evaluated about the COG. In this frame, the inertia tensor is diagonal and remains constant throughout the motion.

The angular velocity vector is decomposed as
\begin{equation}
\vec{\omega} = \omega_1 \vec{e}_1 + \omega_2 \vec{e}_2 + \omega_3 \vec{e}_3 ,
\qquad
\omega_i = \vec{\omega}\cdot\vec{e}_i .
\end{equation}
Similarly, the torque acting on the rigid body is projected onto the principal directions,
\begin{equation}
M_{A_i} = \vec{M}_A \cdot \vec{e}_i .
\end{equation}

The angular momentum balance in the body-fixed frame yields the Euler equations
\begin{align}
M_{A_1} &= I_{11}^A \dot{\omega}_1 - (I_{22}^A - I_{33}^A)\omega_2 \omega_3, \\
M_{A_2} &= I_{22}^A \dot{\omega}_2 - (I_{33}^A - I_{11}^A)\omega_3 \omega_1, \\
M_{A_3} &= I_{33}^A \dot{\omega}_3 - (I_{11}^A - I_{22}^A)\omega_1 \omega_2 .
\end{align}
Solving this system provides the angular acceleration components $\dot{\omega}_i$, which are assembled into
\begin{equation}
\dot{\vec{\omega}} =
\dot{\omega}_1 \vec{e}_1 +
\dot{\omega}_2 \vec{e}_2 +
\dot{\omega}_3 \vec{e}_3 .
\end{equation}

Though the present geometry is a sphere, the above calculations are applicable to a body of complex shape as well. 

\subsubsection{Time integration and rigid-body kinematics}
\label{sec:time}
The translational and rotational motions of the rigid body are advanced in time using a first-order explicit integration scheme,
\begin{equation}
\vec{v}_A^{n+1} = \vec{v}_A^{n} + \Delta t\,\vec{a}_A ,
\qquad
\vec{\omega}^{n+1} = \vec{\omega}^{n} + \Delta t\,\dot{\vec{\omega}} .
\end{equation}
The first-order explicit scheme employed here does not exactly conserve the total kinetic energy of the rigid body. However, since the present simulations focus on short-term dynamics and are not sensitive to small numerical energy drift, this approximation is acceptable. For a simulation involving many particles higher order scheme may be preferred.

The velocity of any material point $P$ belonging to the rigid body is obtained from the rigid body kinematic relation
\begin{equation}
\vec{v}_P = \vec{v}_A + \vec{\omega} \times \vec{r}_{PA} ,
\end{equation}
and its position is updated as
\begin{equation}
\vec{r}_P^{n+1} = \vec{r}_P^{n} + \vec{v}_P \Delta t .
\end{equation}

The principal directions are advanced accordingly
\begin{equation}
\vec{e}_i^{\,n+1}
=
\vec{e}_i^{\,n}
+
\Delta t
\left(
\vec{\omega}^{\,n+1} \times \vec{e}_i^{\,n}
\right) ,
\end{equation}
after which, the basis vectors are normalized to preserve orthonormality and prevent numerical drift.

\subsubsection{Collision detection and rebound modelling}
\label{sec:collision}
When the bead approaches the solid substrate, its vertical motion must be corrected to capture the exact instant of impact and the subsequent rebound. Collision is detected when the lower boundary of the bead, represented by an effective radius $R$, crosses the plane of the solid wall located at $z=z_{\mathrm{wall}}$ within a timestep.

If the bead is moving downward and a collision is detected, the exact impact time is computed by linear interpolation of the COG trajectory. The post-impact vertical velocity is prescribed using the dry coefficient of restitution, i.e. $e_r$,
\begin{equation}
v_z^{+} = -\mathrm{e_r}\, v_z^{-} ,
\end{equation}
The value of $e_r$ is taken from the dry-collision measurements reported by Gollwitzer et al.~(0.976–0.984)~\cite{Gollwitzer}. In the present simulations, we use $e_r$~=~0.98. The bead is advanced through the remaining fraction of the timestep using the rebound velocity. 

The above rebound model represents an idealized, instantaneous rigid-body collision with a prescribed, velocity-independent dry restitution coefficient. This simplified treatment neglects finite contact duration and elastic deformation effects. For the bead–substrate system considered here, the experimentally measured dry restitution reported by Gollwitzer et al.~\cite{Gollwitzer} is close to unity and exhibits only weak variation over the relevant impact velocities, indicating that solid-contact dissipation is small. Since the present study focuses on additional energy losses induced by the liquid layer, this simplified dry rebound model provides a controlled baseline for isolating liquid effects and is not expected to influence the observed wet-impact trends.

After the collision-induced correction of the COG trajectory, a corresponding displacement is applied to all rigid-body Lagrangian particles. 

Near the solid substrate, the bead experiences rapid changes in momentum during approach, impact, and rebound. To resolve these dynamics accurately and maintain numerical stability, the timestep is adaptively reduced when the clearance between the bead’s lower boundary and the wall becomes comparable to the SPH particle spacing. Once the bead moves away from the wall, the timestep is restored to its nominal value.

Together, this combination of numerics provides a physically consistent description of rigid-body translation, rotation, and bead–wall interaction within the updated Lagrangian SPH framework employed in the present study.
Algorithmic details of the collision detection, sub-timestep impact resolution, rigid-body particle shifting, and adaptive timestep strategy are provided in Appendix~A~(Algorithms A1–A3).

\section{Simulation Setup}
\label{sec:setup}
\begin{figure}[ht]
    \centering
    \includegraphics[width=0.65\linewidth]{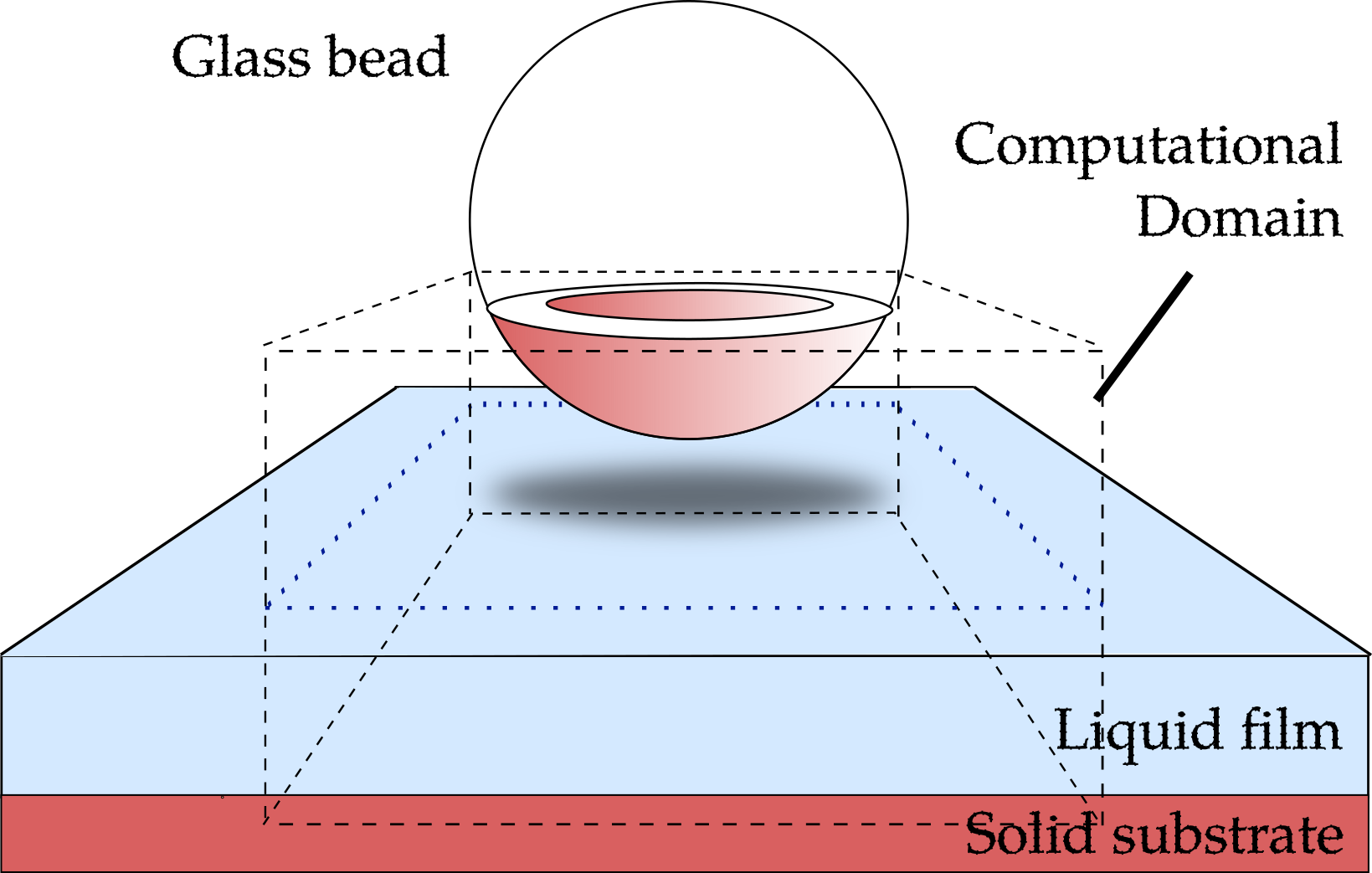}
    \caption{
     The physical scenario and the computational domain}
    \label{fig:setup}
\end{figure}

\begin{table}[htb]
\centering
\caption{Simulation parameters used to analyze size-dependent wet-impact behavior.}
\label{tab:sim_params}
\begin{tabular}{l c c}
\hline
\textbf{Parameter} & \textbf{Symbol} & \textbf{Value / Range} \\
\hline
Impact velocity        & $v$ & 0.10 -- 1.50 m/s \\
Glass bead diameter      & $d$ & 2.8 -- 8.0 mm    \\
Liquid film thickness  & $\delta$ & 0.40, 0.45 mm    \\
\hline
\end{tabular}
\end{table}

As mentioned before, we reduce computational cost of the simulation (without loss in accuracy) by employing a truncated spherical shell to represent the glass bead. Only the lower portion of the bead interacts with the liquid film and the substrate. 
The bead’s rigid-body motion, however, is computed using the physical properties of the full sphere. The translational and rotational dynamics are updated using the full spherical mass, centre of gravity (COG), and moment of inertia, and the motion tracking is performed with respect to the COG of the complete sphere. 
Thus, the model captures the rebound kinematics and fluid–solid interaction behavior without the expense of resolving the full three-dimensional particle.

In the simulations, The bead was released from a height of 0.60 mm and impacted the surface with velocities ranging from approximately 0.1 to 1.5 m/s. The beads have a density of 2.58 g/$cm^3$, and their diameters vary from 2.8 mm to 8 mm to analyze size-dependent wet-impact behavior, as shown in Table~\ref{tab:sim_params}. The target is modeled as a thick, rigid solid substrate to eliminate potential influences from substrate deformation or vibrations. Fig.~\ref{fig:setup} shows the complete numerical setup consisting of the truncated spherical bead, the liquid film, and the rigid substrate. This numerical configuration provides a controlled framework to investigate how liquid properties, film thickness, bead size, and impact velocity collectively influence the COR of wet collisions.

The material properties used in the simulations, which enter the definition of the Stokes number and govern the liquid–solid interaction, are summarized in Table~\ref{tab:material_properties}.
\begin{table}[h!]
\centering
\caption{Material properties used in the simulations. }
\begin{tabular}{lccc}
\toprule
\multirow{2}{1.25in}{Material} & Density  & Dynamic viscosity  & Surface tension  \\
& $\rho$ (kg/m$^3$) & $\eta$ (Pa$\cdot$s) & $\sigma$ (N/m)\\
\midrule
Water          & 998  & 0.001 & 0.0728 \\
M5 silicone oil            & 925  & 0.0046 & 0.0192 \\
Glass bead                 & 2580 & --                 & --    \\
\bottomrule
\end{tabular}
\label{tab:material_properties}
\end{table}

\subsection{Calculation of Coefficient of Restitution}
The coefficient of restitution (COR) is commonly defined as a measure of the kinetic energy recovered after a collision and may be expressed either in terms of mechanical energy change or, more practically, as the ratio of rebound to impact velocity. In wet impacts, however, the particle interacts with a liquid layer prior to solid contact, making the definition of these velocities less straightforward.

In experimental studies of wet particle impacts, the velocity-ratio definition is typically employed with velocities determined outside the liquid layer, for example, from trajectory measurements before entry into and after exit from the film (\cite{Joseph2001}, \cite{Gollwitzer}). In this sense, the reported velocities correspond to the particle speed immediately before penetrating and immediately after emerging from the liquid layer.

In the present simulations, we adopt the same conceptual definition while exploiting the precise geometric information available numerically. The impact velocity is defined as the vertical velocity of the bead at the instant its lower surface first makes contact with the initial, undisturbed liquid–air interface. After penetrating the film, colliding with the solid substrate, and rebounding upward, the rebound velocity is evaluated when the bead surface re-emerges through this same reference height during ascent. Although the interface undergoes transient deformation during impact, using the initial interface position as a fixed geometric reference ensures that both velocities are evaluated at an identical vertical location.

Thus, the wet coefficient of restitution is computed as:
\begin{equation}
COR = \frac{v_{\mathrm{rebound}}}{v_{\mathrm{impact}}}.
\end{equation}

This approach provides a physically consistent and repeatable measure of the coefficient of restitution under wet-impact conditions.

Using the simulation setup and the numerical methodology described in the previous section, we present a validation of the model before venturing into parametric studies. 


\section{Validation}
\label{sec:validation}
We establish the accuracy and robustness of the ISPH model through validation simulations against available experimental data and a resolution study. The ISPH formulation is validated against the experimental measurements of Gollwitzer \textit{et al.}~\cite{Gollwitzer} for the normal impact of a 5.5~mm glass bead onto a 0.45~mm thick M5 silicone-oil film. Simulations were performed over the same range of impact velocities, and rebound velocities were extracted for direct comparison.

To assess spatial convergence and agreement with experiments, simulations were conducted using liquid resolutions between 22,445 and 247,500 particles. The relative $L_2$ error in rebound velocity was computed with respect to the experimental data, taken as reference values, according to
\begin{equation}
\|e\|_{L_2} =
\frac{\sqrt{\sum_i \left( u_i^{\mathrm{num}} - u_i^{\mathrm{exp}} \right)^2}}
{\sqrt{\sum_i \left( u_i^{\mathrm{exp}} \right)^2}},
\end{equation}
where $u_i^{\mathrm{num}}$ and $u_i^{\mathrm{exp}}$ denote the numerical and experimental rebound velocities, respectively.

\begin{figure}[ht]
\centering
\begin{subfigure}{0.53\textwidth}
    \includegraphics[width=\textwidth]{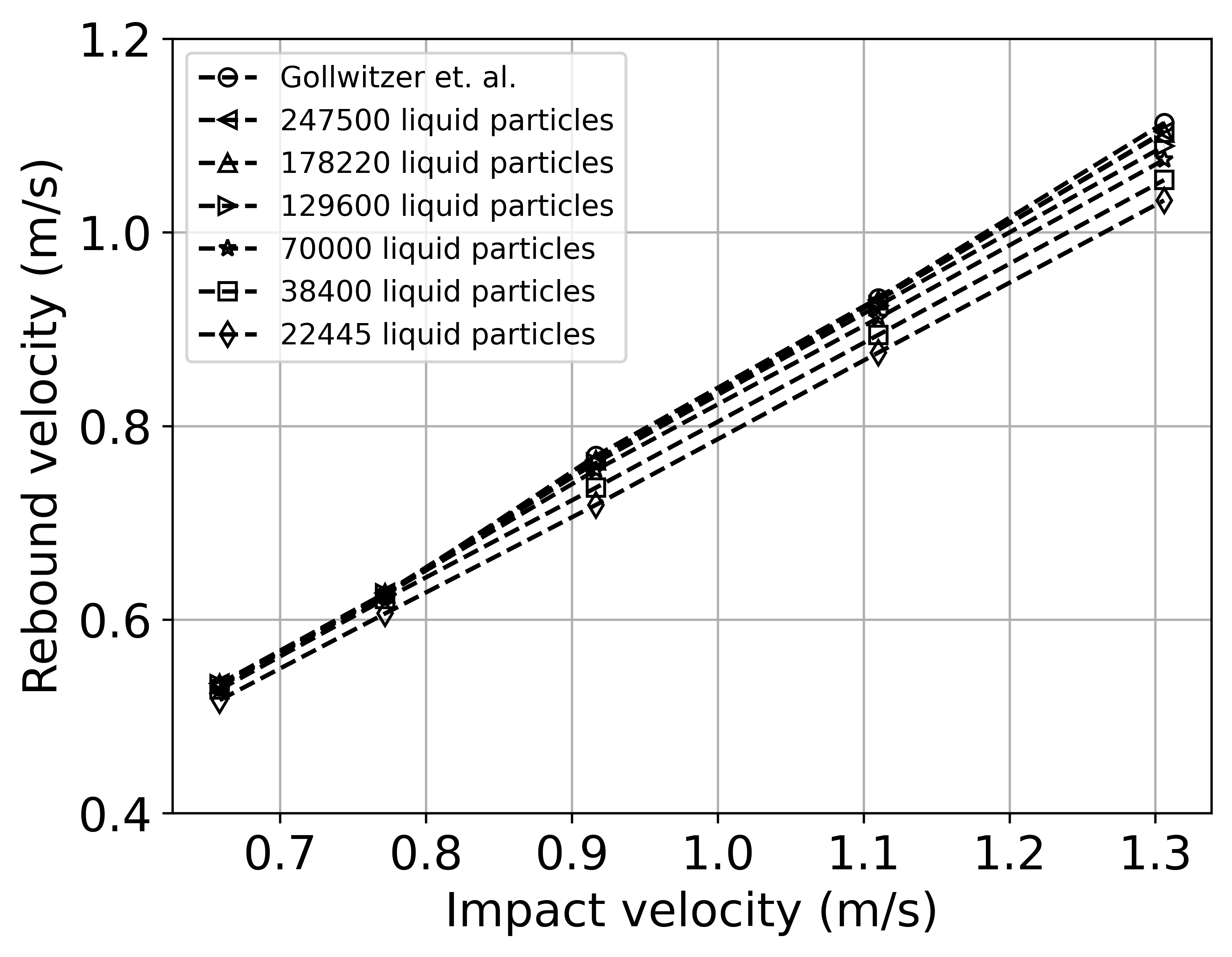}
    \caption{
     }
\end{subfigure}
 \begin{subfigure}{0.45\textwidth}   
    \includegraphics[width=\textwidth]{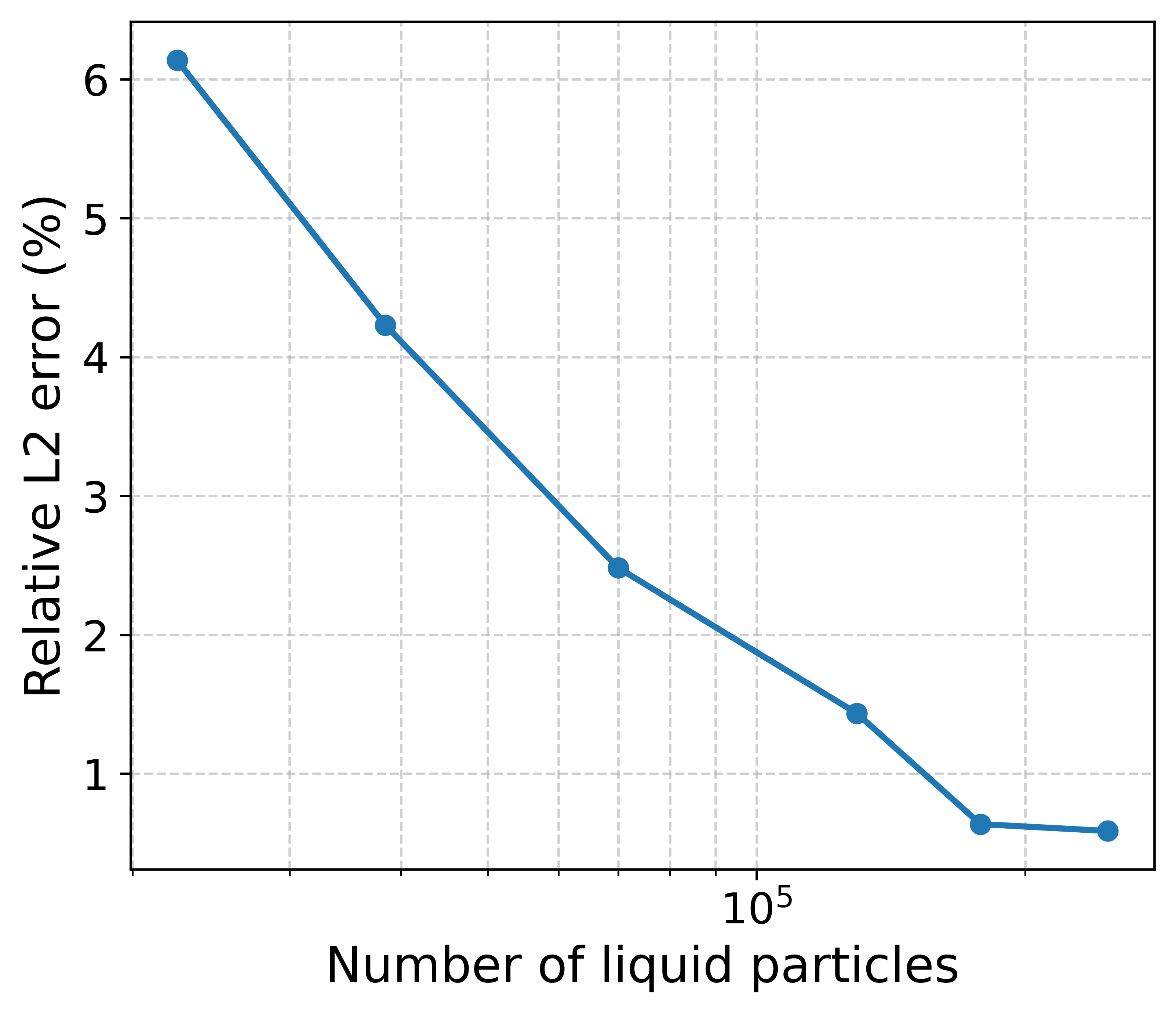}
    \caption{}
    \label{fig:l2_error}
    \end{subfigure}
    \caption{(a)Validation and particle-resolution analysis for the impact of a 5.5 mm glass bead onto a 0.45 mm thick M5 silicone-oil film. The rebound velocity obtained from ISPH simulations is compared with experimental measurements of Gollwitzer et al. \cite{Gollwitzer} for increasing liquid particle resolution. (b) Relative $L_2$ error in rebound velocity as a function of liquid particle number for the impact of a 5.5~mm glass bead onto a 0.45~mm M5 silicone-oil film. The error decreases monotonically with increasing resolution and approaches convergence beyond approximately 178,000 particles. }
    \label{fig:validation}
\end{figure}

We present the rebound velocity calculated from simulations for a number of cases at different particle resolutions in Fig. \ref{fig:validation}. Figure~\ref{fig:l2_error} shows the resulting error as a function of liquid particle resolution. The error decreases systematically with increasing resolution, from approximately 6\% at the coarsest discretization to below 1\% for the two finest cases. The reduction becomes small beyond 178,220 liquid particles (0.637\% at 178,220 particles compared to 0.588\% at 247,500 liquid particles), indicating that the solution is effectively converged with respect to experimental measurements.

\begin{figure}[ht]
    \centering
    \includegraphics[width=3.5in]{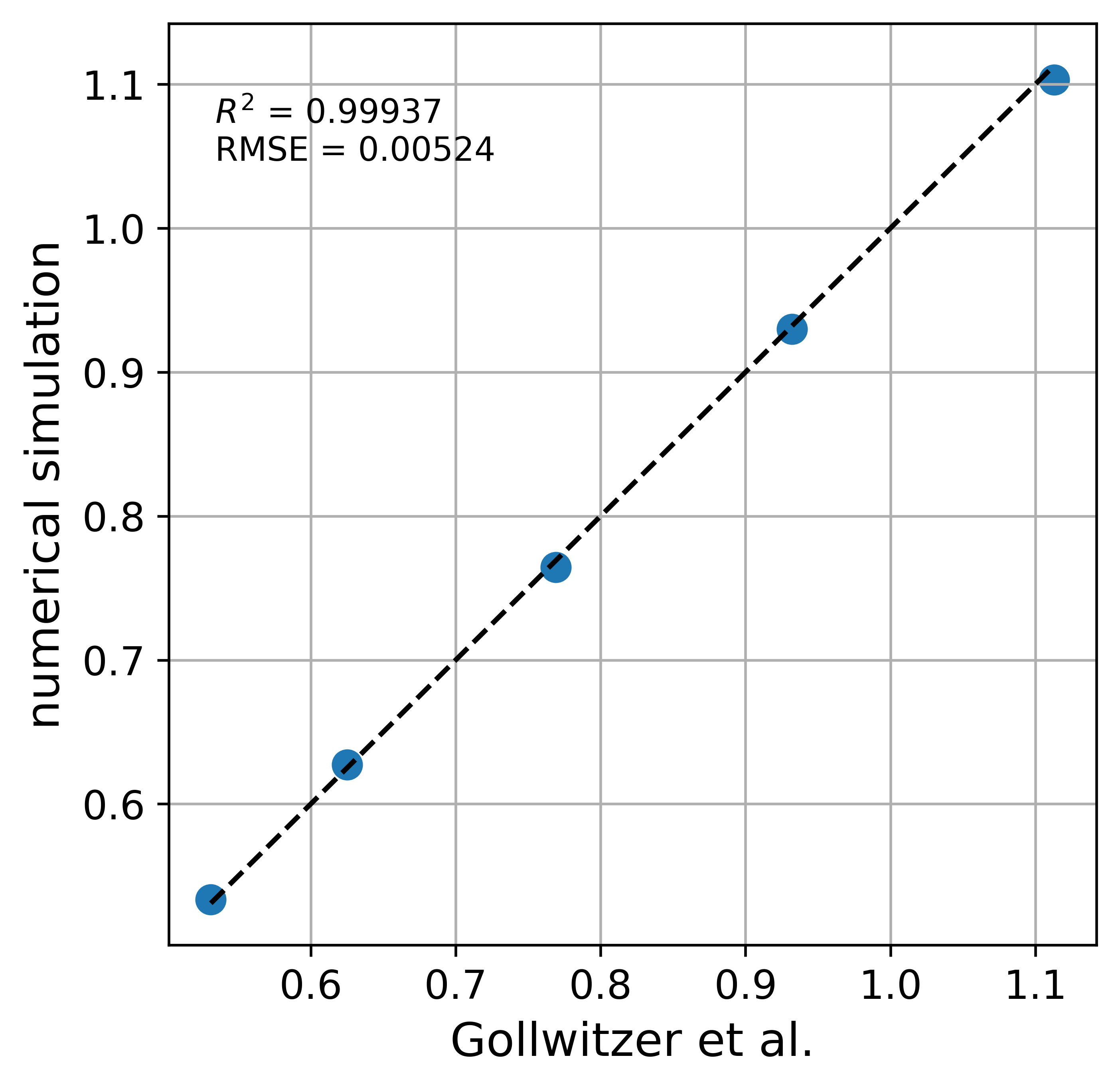}
\caption{Parity plot comparing simulated and experimental \cite{Gollwitzer} rebound velocities for the 178,220-particle resolution. The dashed line represents perfect agreement. The data cluster tightly around the identity line ($R^2 = 0.99937$, $\mathrm{RMSE} = 0.00524$), indicating excellent agreement with experiments.}

    \label{fig:parity_plot}
\end{figure}

The 178,220 liquid particle resolution was therefore adopted for subsequent simulations, as it provides nearly identical accuracy at a substantially lower computational cost. At this resolution, each simulation requires approximately 12~hours on a 20-core Intel i9 processor. Figure~\ref{fig:parity_plot} presents a parity comparison between simulated and experimental rebound velocities for the selected resolution. The data lie close to the identity line, with $R^2 = 0.99937$ and $\mathrm{RMSE} = 0.00524$, confirming the consistency between simulation and experiment across the examined velocity range.
\section{Results and Discussions}
\label{sec:results}
\begin{figure}[ht]
    \centering
    \includegraphics[width=6in]{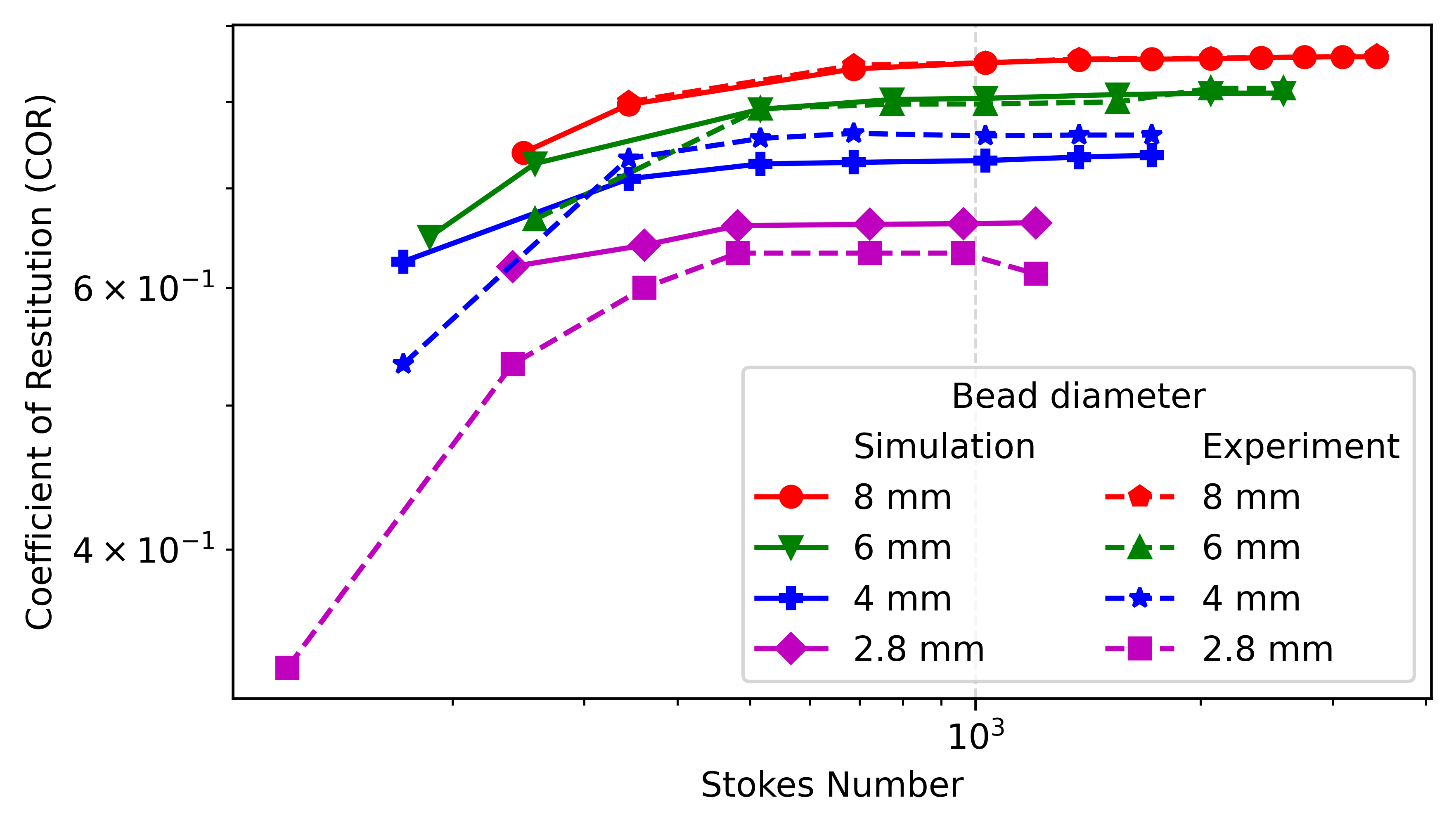}
    \caption{
Coefficient of restitution (COR) as a function of Stokes number (St) for different liquid thicknesses of Water, compared with the experimental results of Gollwitzer et. al. \cite{Gollwitzer}. 
Evidently, the Stokes number doesn't fully characterize wet-impact rebound. 
    }
    \label{fig:comparison}
\end{figure}

Here, we present a parameter study and subsequent analysis on the rebound of a rigid sphere on a plane wet substrate.
We use the Stokes number defined in Gollwitzer \textit{et al.} \cite{Gollwitzer} as
\begin{equation}
St = \frac{\rho_g D v_{\mathrm{impact}}}{9\eta},
\end{equation}
which compares particle inertia to viscous resistance in the liquid film, where $\rho_g$ is the glass bead density, $D$ is the glass bead diameter, $v_{\mathrm{impact}}$ is the impact velocity, and $\eta$ is the dynamic viscosity of the liquid. In \cite{Gollwitzer}, the COR is shown to scale with the Stokes number alone, which was demonstrated with a relatively thin liquid film. As the film becomes thicker relative to the bead size, we postulate that the flow dynamics in the film start to play a role. To substantiate this, we introduce an additional nondimensional quantity that we call the dimensionless film thickness, defined as
\begin{equation}
\gamma = \frac{\delta}{D},
\end{equation}
where $\delta$ is the film thickness and $D$ is the bead diameter. 
For the impact conditions considered in this study, the Weber number~(
$\mathrm{We} = (\rho_l v_{\mathrm{impact}}^{2} D)/\sigma
$ ),
where $\rho_l$ and $\sigma$ denote the liquid density and liquid surface tension, respectively) lies in the range $\mathcal{O}(10)$–$\mathcal{O}(10^2)$, indicating a moderate-to-high Weber number regime in which inertial effects dominate over surface tension forces. We therefore neglect surface tension effects for this study. All results presented in this section are obtained using the liquid particle resolution validated in \S \ref{sec:validation}. We maintain the same particle resolution relative to the film thickness across all bead diameters and liquid configurations in this section.

The simulation results are compared with the experimental data from \cite{Gollwitzer} in Fig.~\ref{fig:comparison}. For the largest diameter (8~mm), the agreement is very good across the range of \Sto. However, deviations appear for the smaller diameters (6, 4, and 2.8~mm), even at comparable \Sto. As the particle diameter decreases, the corresponding Weber number also decreases, and capillary effects become more significant in the experiments, whereas surface tension is not included in the present model.

Independently of the comparison with experiments, Fig.~\ref{fig:comparison} shows that the wet coefficient of restitution cannot be expressed as a power-law monomial of \Sto~alone (see \cite{barenblattscaling}). At fixed \Sto, the COR varies systematically with particle diameter, with larger particles exhibiting consistently higher rebounds. The curves do not collapse onto a single trend when plotted against \Sto. This systematic separation indicates that a single-parameter scaling in \Sto~is insufficient to characterize the rebound dynamics, motivating the introduction of a non-dimensional thickness parameter.

\begin{figure}[ht]
    \centering
    \includegraphics[width=3.5in]{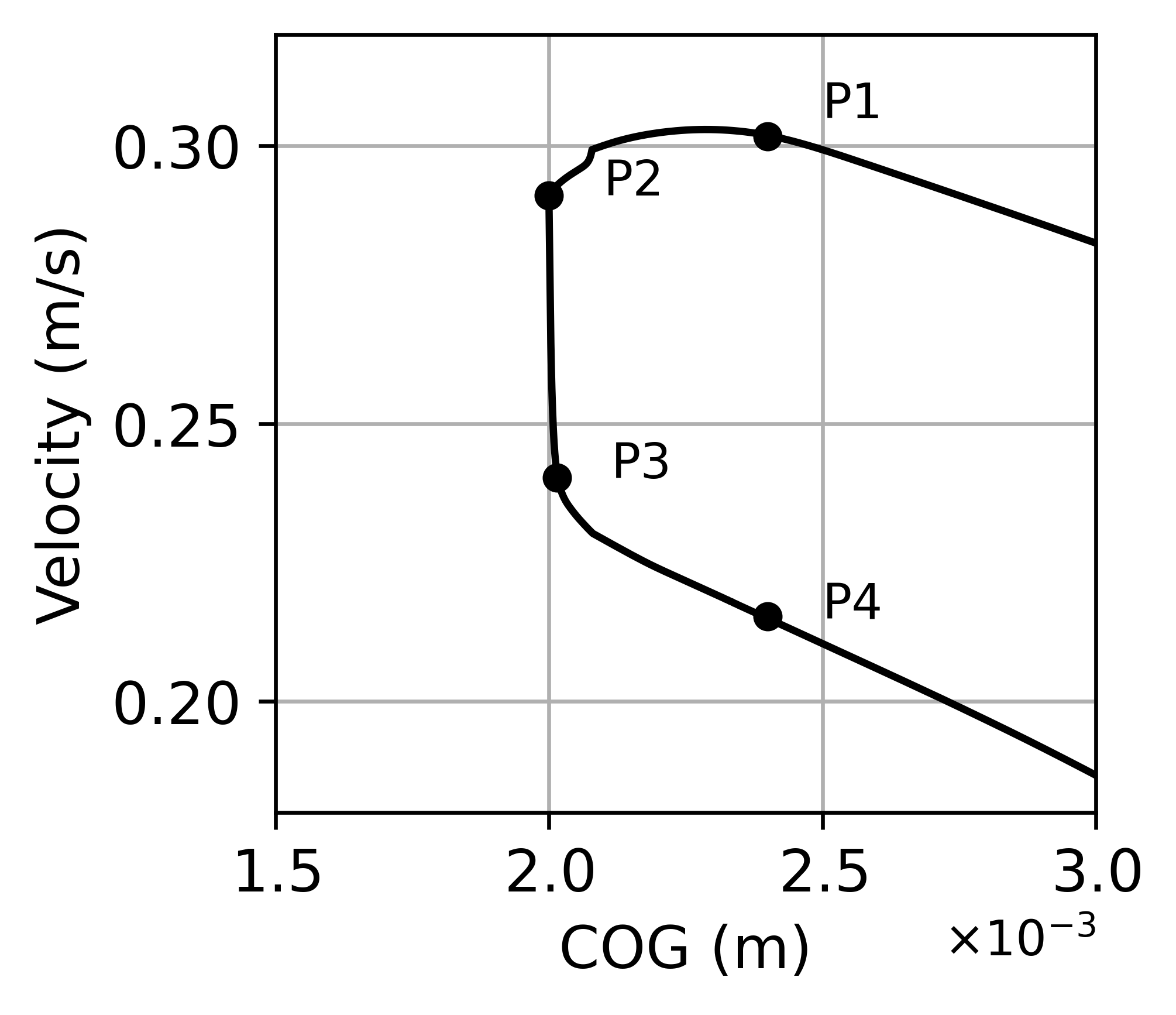}
    \caption{
     Position of center of gravity (COG) of the sphere vs. its normal velocity (absolute value) during impact onto a thin liquid film at an impact velocity of 0.30 m/s. The labeled points P1–P4 denote key stages of the interaction: initial contact with the liquid film (P1), collision with the solid substrate (P2 \& P3) and separation from the liquid film defining the rebound velocity (P4).}
    \label{fig:velocityp3}
\end{figure}

\begin{figure}[ht]
    \centering
    \begin{subfigure}[b]{0.48\linewidth}
        \centering
        \includegraphics[width=\textwidth]{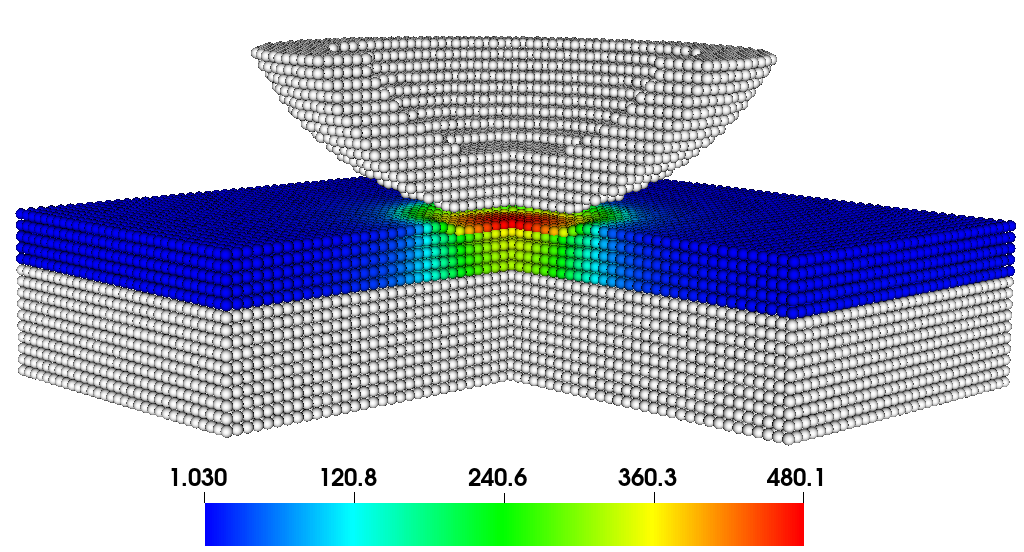}
        \caption{at P1 in Figure 4}
        \label{fig:sub1}
    \end{subfigure}
    \hfill
    \begin{subfigure}[b]{0.48\linewidth}
        \centering
        \includegraphics[width=\textwidth]{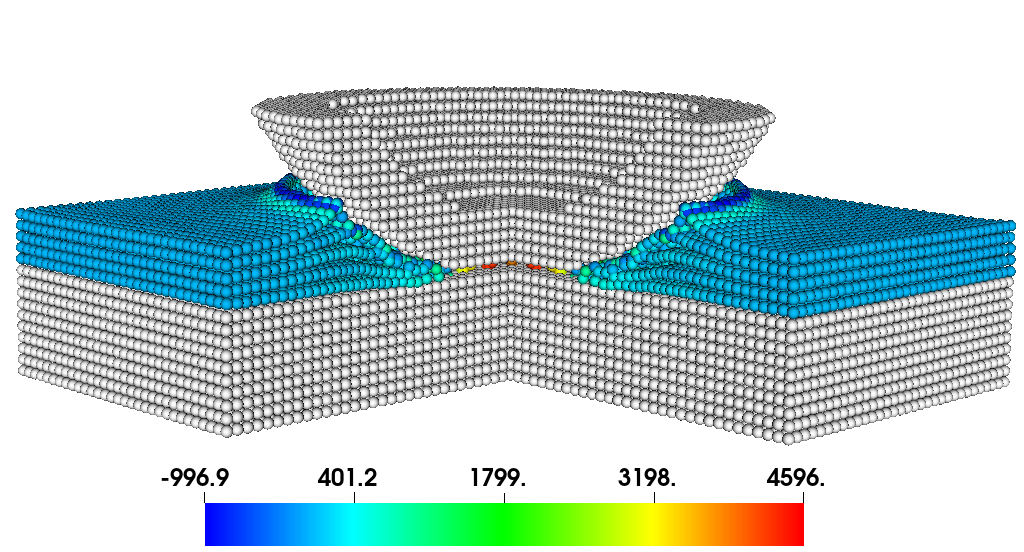}
        \caption{at P2 in Figure 4}
        \label{fig:sub2}
    \end{subfigure}
    \vskip\baselineskip
    \begin{subfigure}[b]{0.48\linewidth}
        \centering
        \includegraphics[width=\textwidth]{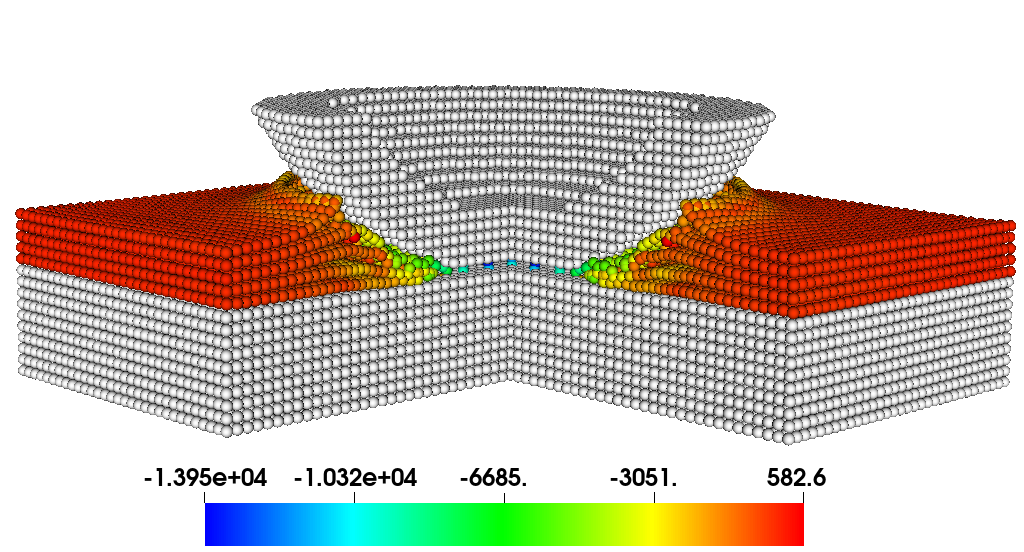}
        \caption{at P3 in Figure 4}
        \label{fig:sub3}
    \end{subfigure}
    \hfill
    \begin{subfigure}[b]{0.48\linewidth}
        \centering
        \includegraphics[width=\textwidth]{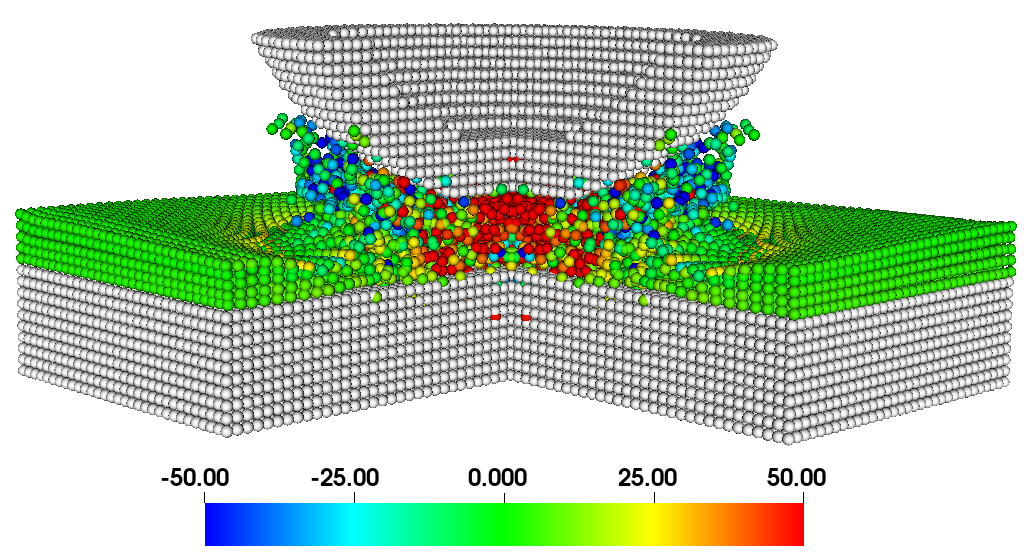}
        \caption{at P4 in Figure 4}
        \label{fig:sub4}
    \end{subfigure}
    \caption{Evolution of the pressure field in the liquid film during impact at an impact velocity of 0.30 m/s. The pressure buildup beneath the sphere intensifies after substrate contact (P2), providing resistance to the motion of the rigid body, and subsequently relaxes as the sphere approaches separation from the liquid film. The color map reports gauge pressure (same units as in the scale bars).}
    \label{fig:four_subfig}
\end{figure}

We perform several simulations of the normal impact of a rigid sphere on a layer of water. The various cases considered are tabulated in Table \ref{tab:sim_params}. As an exemplary simulation, we show the evolution of the absolute value of the normal velocity of the rigid sphere during its impact on a 0.4 mm water film at an impact velocity of 0.30 m/s in Fig.~\ref{fig:velocityp3}. Prior to contact with the liquid film (denoted by the phase-space trajectory before \emph{P1} indicated in Fig. \ref{fig:velocityp3}), the sphere undergoes free fall under gravity, resulting in a linear increase in velocity with time. Upon contact with the liquid film at the instance \emph{P1}, hydrodynamic forces begin to act on the sphere, modifying its motion. Between \emph{P1} and \emph{P2}, the velocity continues to increase in spite of viscous resistance for a duration after which it decreases until \emph{P2}. The point \emph{P2} corresponds to the collision of the sphere with the solid substrate beneath the liquid film. Following substrate contact, the dry coefficient of restitution is affected, resulting in an abrupt loss in the  velocity indicated by the phase-space trajectory from \emph{P2} to \emph{P3}. Between\emph{P3} and \emph{P4}, the deceleration is governed by the combined effects of viscous drag and form drag arising from fluid interaction. The pressure fields in the liquid film during these instances in the phase-space trajectory are also presented in Fig. \ref{fig:four_subfig}. Specifically, in Fig. \ref{fig:four_subfig}(c)~(corresponding to \emph{P3}) a suction develops beneath the sphere after substrate contact, opposing its motion. As the suction pressure relaxes and the hydrodynamic forces weaken, the sphere approaches separation from the liquid film. At P4 (at the same vertical position as \emph{P1}), the sphere detaches from the liquid layer, and the velocity at this instant is identified as the rebound velocity used to compute the wet coefficient of restitution.

\begin{figure}[ht]
    \centering
    \begin{overpic}[percent,width=0.75\textwidth]{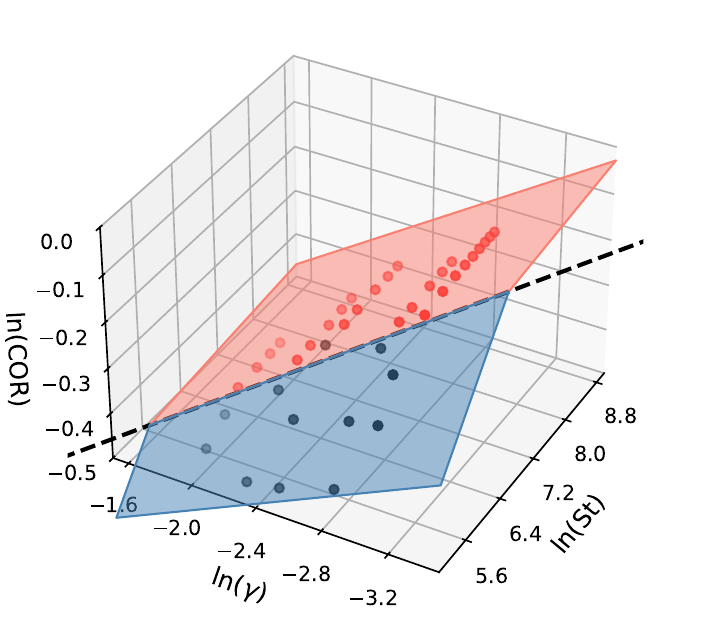}
        
       \put(35, 25){\Large \texttt{R1}} 
        
        \put(60, 55){\Large \texttt{R2}}
        \put(80, 52){\Large $\mathcal{L}$}
    \end{overpic}
    \caption{Wet coefficient of restitution (COR) as a function of the Stokes number $St$ and the dimensionless film thickness $\gamma = \delta/D$. 
    Two regimes, R1 and R2, are evident, represented by linear regression planes in blue and orange respectively. The planes intersect along the line $\mathcal{L}$.} 
\label{fig:stokes_3d}
    
    \end{figure}
    
\begin{figure}[ht]
\centering
\includegraphics[width=0.75\textwidth]{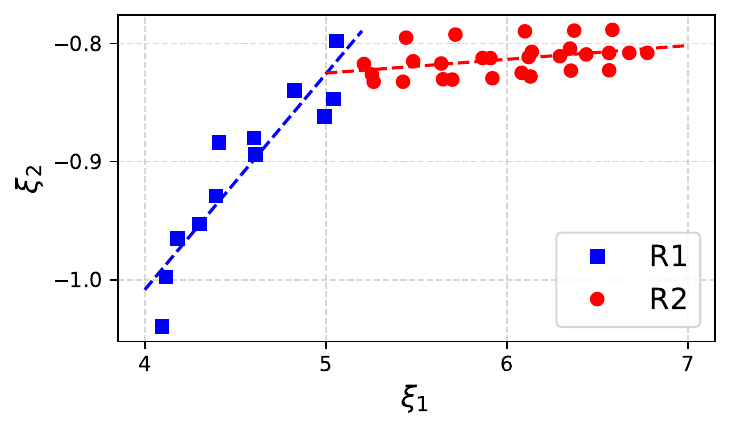}
    \caption{The COR is implicitly plotted using transformed coordinates: $\xi_1= 0.95\ln{St} + 0.31\ln{\gamma} -0.07\ln{COR}$ and $\xi_2 = 0.22\ln{\gamma}+0.97\ln{COR}$, obtained through Gram-Schmidt orthogonalization from the basis of Fig.~\ref{fig:stokes_3d}. The dash lines are the corresponding projections of planes \texttt{R1} and \texttt{R2} in Fig.~\ref{fig:stokes_3d}. Two sets of scaling exponents are immediately evident.
    }   
    \label{fig:stokes_2d}
\end{figure}

\subsection{Scaling regimes}

Using our simulations we seek to obtain a power law monomial formula for the \Cor~as a function of \Sto~as well as $\gamma$ . That is, we seek 
\begin{equation}
\Cor \sim \Sto^{a}\,\gamma^{b}.
\end{equation}
where $a$ and $b$ are power-law exponents characterizing the Stokes-number and geometric contributions, respectively. We adopt the same definition of the Stokes number as Gollwitzer \textit{et al.}~\cite{Gollwitzer}. Taking natural logarithm on both sides, we get 
\begin{equation}
\ln\Cor \sim a\ln\Sto + b\ln\gamma.
\end{equation}
Therefore we present the variation of $\ln$\Cor~with  $\ln$\Sto~and  $\ln\gamma$ in Fig. \ref{fig:stokes_3d}. In the most cases considered in this figure, the liquid-film thickness is held fixed at $\delta=0.40$ mm while the particle diameter D is varied, except for one additional case at D=5.5 mm, which  uses $\delta$ =0.45~mm to probe the sensitivity to $\gamma$ at a comparable Stokes number.  At a given Stokes number, the wet coefficient of restitution decreases with increasing $\gamma$ everywhere. Thus form factor clearly plays a role in determining \Cor. 

Two distinct scaling regimes are evident from Fig.~\ref{fig:stokes_3d}. To quantify these trends, we perform a linear least-squares regression in logarithmic space separately for each regime. The resulting fitted planes, marked as \texttt{R1} and \texttt{R2} in Fig.~\ref{fig:stokes_3d} (blue and orange, respectively), correspond to the following relations:

\begin{align}
\text{For \texttt{R1}: } \ln\Cor &= -1.76  +  0.17 \ln\Sto  -0.16 \ln\gamma  \label{eq:r1}\\
\text{For \texttt{R2}: } \ln\Cor &=  -0.90 +  0.01 \ln\Sto   -0.22 \ln\gamma  \label{eq:r2}
\end{align}

The regression quality was quantified in logarithmic space. 
For regime \texttt{R1}, the fit yields $R^2 = 0.915$ with a log-space root-mean-square error (RMSE) of $0.026$. 
Both fitted exponents are statistically significant ($p < 10^{-4}$), indicating a robust dependence of \Cor~on both \Sto~and $\gamma$ in this regime. 

For regime \texttt{R2}, the regression yields $R^2 = 0.979$ and RMSE = $0.013$. The $\gamma$ exponent remains highly significant ($p \ll 0.01$), whereas the \Sto~exponent is only marginally significant ($p = 0.053$), consistent with the very weak \Sto~dependence observed in Eq.~(\ref{eq:r2}). Taken together, these results indicate that in \texttt{R1} both inertial and geometric effects influence restitution, whereas in \texttt{R2} the dynamics are governed primarily by $\gamma$.

The two fitted planes intersect along a straight line $\mathcal{L}$ in logarithmic space, which represents the boundary separating the two scaling regimes. This intersection line can be parametrically expressed in the $\ln\Sto$–$\ln\gamma$–$\ln\Cor$ coordinates as:

\begin{equation}
\mathcal{L} = ( 5.22+ 0.32t,  -0.92t, -0.86+0.2t),
\end{equation}
where $t$ is an independent parameter.

The corresponding piecewise scaling law in physical variables, therefore, becomes
\begin{equation}
\Cor =
\begin{cases}
0.17 \Sto^{0.17}  \gamma^{-0.16} & \text{if } s<0 \text{ (regime \texttt{R1})},\\
0.4 \Sto^{0.01}\gamma^{-0.22} & \text{otherwise (regime \texttt{R2})},
\end{cases}
\end{equation}
where $s = 0.16\ln \Sto + 0.06\ln\gamma - 0.85$, obtained by subtracting Eq.~(\ref{eq:r2}) from Eq.~(\ref{eq:r1}).

In Fig. \ref{fig:stokes_3d}, the fact that the data are organized along the planes may not be very evident. Hence, we provide in Fig. \ref{fig:stokes_2d} the same data transformed in such a way that the line of intersection of planes \texttt{R1} and \texttt{R2} in Fig. \ref{fig:stokes_3d} is perpendicular to the plane of the paper. We achieved this with a simple Gram-Schmidt orthogonalization on the coordinates of Fig.~\ref{fig:stokes_3d}, by choosing the coordinate along $\mathcal{L}$ as a starting direction. We get $\xi_1= 0.95\ln{St} + 0.31\ln{\gamma} -0.07\ln{COR}$ and $\xi_2 = 0.22\ln{\gamma}+0.97\ln{COR}$ as a result of this orthogonalization. The fitted planes are also transformed, and they show up as dotted lines in Fig. \ref{fig:stokes_2d}. The two regime scaling is evident.

The existence of two collision regimes suggests that different physical phenomena may be at play. Since the focus here is on the restitution coefficient, an attempt to explain the effect of the exact flow phenomena is beyond the scope of the present work. We therefore provide a simple and qualitative explanation based on the total kinetic energy dissipated through the liquid film. 

\begin{figure}[ht]
    \centering
    \begin{subfigure}[b]{0.48\linewidth}
        \centering
        \includegraphics[width=\textwidth]{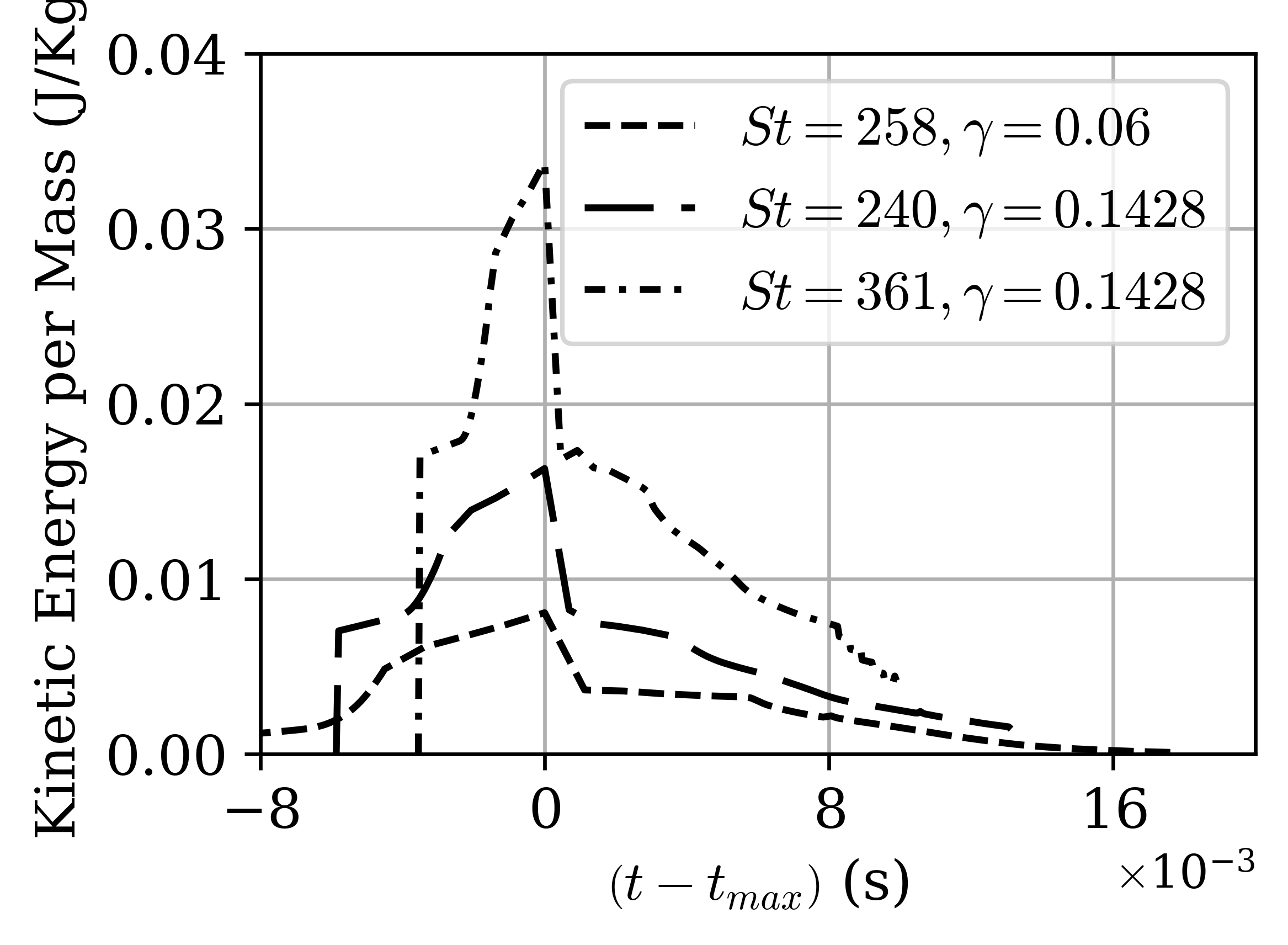}
        \caption{for regime R1}
        \label{fig:ke1}
    \end{subfigure}
    \hfill
    \begin{subfigure}[b]{0.48\linewidth}
        \centering
        \includegraphics[width=\textwidth]{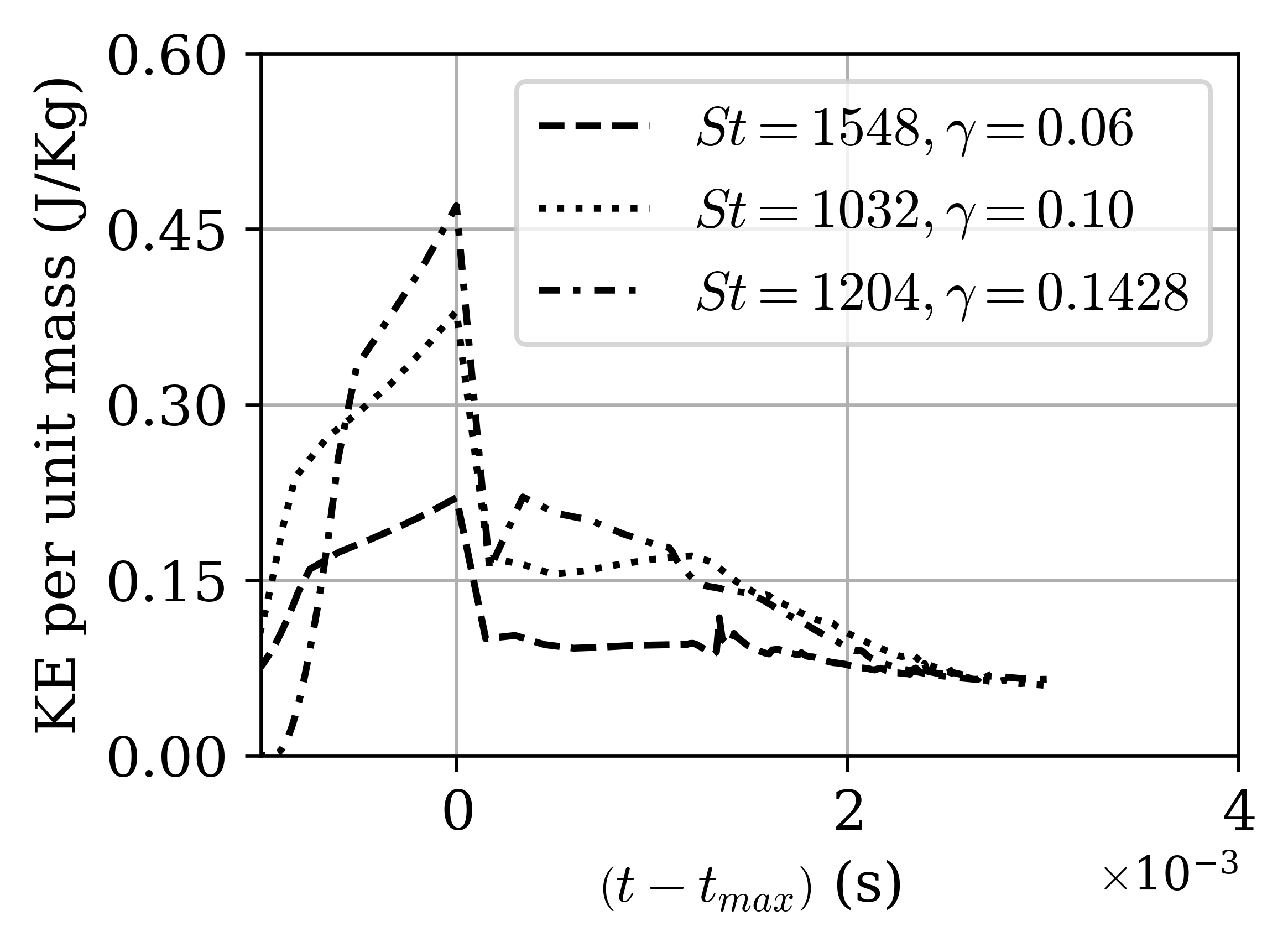}
        \caption{for regime R2}
        \label{fig:ke2}
    \end{subfigure}
\caption{Liquid-phase kinetic energy per unit mass as a function of time, for both the regimes (a) R1 and (b) R2. Time axis is shifted by $t_{\text{max}},$ the instance when the kinetic energy per unit mass in the liquid is the maximum. While (a) shows monotonic decrease in kinetic energy per unit mass after $t_\text{max}$, all the cases in (b) show a spike indicating secondary flow phenomena.}
    \label{fig:KE}
\end{figure}

In Fig.~\ref{fig:KE}, we show the time evolution of the total kinetic energy (KE) in the fluid during the collision process. We randomly pick 3 cases from each of the regimes and plot the evolution of KE in the fluid in each of Figs.~\ref{fig:ke1} and ~\ref{fig:ke2}, respectively. The first peak in each of the curves represents the instance when the bead touches the bottom wall. Until the bead reaches the wall, it continues to impart KE to the fluid. Thereafter, as the bead leaves, we expect the KE in the fluid to be dissipated due to secondary flow phenomena, viscosity, as well as surface waves. In Fig.~\ref{fig:ke1} (corresponding to \texttt{R1}), after the bead contacts the bottom wall, the total KE monotonically dissipates with time. In contrast, in Fig.~\ref{fig:ke2} (corresponding to \texttt{R2}), the total KE does not decrease monotonically. A clear second peak is observed in each of the cases. This indicates that in \texttt{R2}, where the scaling of \Cor~with~\Sto~is weak (as discussed before), secondary flow phenomena such as vortices play a larger role in the dissipation mechanism.  In contrast, the absence of such peaks in Fig. \ref{fig:ke1} indicates a primary flow scenario that only affects the viscous dissipation of energy and hence a stronger scaling with \Sto. Thus, the difference in energy dissipation has contributed to the existence of two different scaling regimes. 

%
\section{Conclusion}
We investigate the normal impact of a rigid spherical bead onto a thin liquid film in the moderate-to-high Weber number regime using incompressible smoothed particle hydrodynamics simulations with coupled rigid-body dynamics to perform a parameter study and to examine the energy dissipation in the fluid film. We present a rigid body interaction algorithm that considerably saves on the computational cost, but reducing the number of degrees of freedom in the numerical computation. We validate this solver against results in literature. 

The simulations show that KE dissipation during wet impacts arises from the combined effects of viscous resistance and flow phenomena within the confined liquid layer. The rebound process proceeds through distinct stages: initial contact with the liquid film, collision with the solid substrate, hydrodynamic loading during confinement, and eventual separation from the film, which together determine the wet coefficient of restitution.

A systematic parametric study demonstrates that the Stokes number alone is insufficient to characterize the wet coefficient of restitution when the liquid-film thickness and particle diameter vary independently. Introducing a dimensionless film thickness, defined as the ratio of the liquid-film thickness to the particle diameter, enables a consistent description of the numerical data. Two distinct regimes are identified based on the dimensionless film thickness. For low values of the dimensionless film thickness, corresponding to larger particles, the coefficient of restitution exhibits a strong dependence on both the Stokes number and the dimensionless film thickness. In contrast, for higher values of the dimensionless film thickness, particularly in the regime we identify as \texttt{R2} corresponding to smaller particles, the overall dependence on the Stokes number is weak. 


We provide a power-law monomial with two sets of exponents for these regimes. This approach may be extended to other fluids with a similar analysis. 
 We further postulate that the regimes may be qualitatively explained by the presence of secondary flow effects evident from the kinetic energy evolution in the liquid layer.
 The developed scaling law models the energy losses in interparticle collisions useful in DEM simulations and other statistical simulation approaches.

\section*{Acknowledgements}
The work is supported by the Science Education and Research Board (SERB) through the Core Research Grant (CRG) number CRG/2022/008634, the Startup Research Grant (SRG) number SRG/2022/000436 and the Ministry of Ports, Shipping and Inland Waterways, India.

\appendix
\section{Algorithms for rigid-body collision handling}

This appendix summarizes the numerical procedures used to implement collision detection, rebound, rigid-body coherence, and timestep adaptation.
\renewcommand{\thealgorithm}{A\arabic{algorithm}}
\begin{algorithm}[h!]
\caption{Sub-timestep collision detection and COG velocity update}
\begin{algorithmic}[1]
\State \textbf{Input:} $z, v_z, R, z_{\text{wall}}, \Delta t, e_r$
\State \textbf{Output:} $z_{\text{new}}, v_{z,\text{new}}$

\State Predict next COG position:
\State $z_{\text{next}} = z + v_z \Delta t$

\If{$v_z < 0$ \textbf{and} $(z_{\text{next}} - R) \le z_{\text{wall}}$}
    \State $\Delta t_{\text{impact}} = \dfrac{(z - R) - z_{\text{wall}}}{|v_z|}$
    \State $z_{\text{impact}} = z + v_z \Delta t_{\text{impact}}$
    \State $v_{z,\text{rebound}} = -e_r\, v_z$
    \State $\Delta t_{\text{remaining}} = \Delta t - \Delta t_{\text{impact}}$
    \State $z_{\text{new}} = z_{\text{impact}} + v_{z,\text{rebound}} \Delta t_{\text{remaining}}$
    \State $v_{z,\text{new}} = v_{z,\text{rebound}}$
\Else
    \State $z_{\text{new}} = z_{\text{next}}$
    \State $v_{z,\text{new}} = v_z$
\EndIf
\end{algorithmic}
\end{algorithm}

\begin{algorithm}[h!]
\caption{Rigid-body particle shifting}
\begin{algorithmic}[1]
\State \textbf{Input:} rigid-body particles $\{p\}$, COG displacement $\Delta z$

\For{each particle $p$}
    \State $p.z \leftarrow p.z + \Delta z$
\EndFor
\end{algorithmic}
\end{algorithm}

\begin{algorithm}[h!]
\caption{Adaptive timestep selection near the wall}
\begin{algorithmic}[1]
\State \textbf{Input:} $z, R, z_{\text{wall}}, \Delta t_{\text{nominal}}, \text{particle\_spacing}$
\State \textbf{Output:} $\Delta t$

\State $\text{clearance} = (z - R) - z_{\text{wall}}$

\If{$\text{clearance} < \text{particle\_spacing}$}
    \State $\Delta t = 0.1\, \Delta t_{\text{nominal}}$
\Else
    \State $\Delta t = \Delta t_{\text{nominal}}$
\EndIf
\end{algorithmic}
\end{algorithm}
\pagebreak
\newpage




\clearpage
\bibliographystyle{elsarticle-num-names} 

\end{document}